\documentclass[preprint]{aastex}
\usepackage{epstopdf}
\usepackage{float}
\usepackage{pdflscape}
\usepackage{subfigure}
\usepackage{lineno}

\usepackage{graphicx, subfigure}
\usepackage{float}
\graphicspath{{plots/} }

\begin{document}
\title{Modeling the Emission from Turbulent Relativistic Jets in Active Galactic Nuclei}
\author{Victoria Calafut\altaffilmark{1},\altaffilmark{2} and Paul J. Wiita\altaffilmark{1} }
\email{calafuv1@tcnj.edu, wiitap@tcnj.edu} 
\altaffiltext{1}{Department of Physics, The College of New Jersey, 2000 Pennington Road, Ewing, NJ, 08628-0718, USA}
\altaffiltext{2}{Currently at the Department of Astronomy, Cornell University, Space Sciences Building, Ithaca, NY, 14853, USA }

\date{\today}

\begin{abstract}
We present a numerical model developed to calculate observed fluxes of relativistic jets in active galactic nuclei. The observed flux of each turbulent eddy is dependent upon its variable Doppler boosting factor, computed as a function of the relativistic sum of the individual eddy and bulk jet velocities and our viewing angle to the jet. The total observed flux is found by integrating the radiation from the eddies over the turbulent spectrum. We consider jets that contain turbulent eddies that have either standard Kolmogorov or recently derived relativistic turbulence spectra.  We also account for the time delays in receiving the emission of the eddies due to their different simulated positions in the jet, as well as due to the varying beaming directions as they turn over. We examine these theoretical light curves and compute power spectral densities (PSDs) for a range of viewing angles, bulk velocities of the jet, and turbulent velocities. These PSD slopes depend significantly on the turbulent velocity and are essentially independent of viewing angle and bulk velocity. The flux variations produced in the simulations for realistic values of the parameters tested are consistent with the types of variations observed in radio-loud AGN as, for example, recently measured with the Kepler satellite, as long as the turbulent velocities are not too high.
\end{abstract} 

\subjectheadings{galaxies: active -- galaxies: jets -- quasars: general}

\section{Introduction}
An active galactic nucleus (AGN) is a compact region at the center of a galaxy characterized by powerful electromagnetic emission and significant flux variability. The AGN is the central engine, fundamentally comprised of a hot accretion disk surrounding a supermassive black hole. An immense amount of radiation is generated by the gravitational infall of material that is heated in the accretion disk.   AGN such as quasars are almost always variable, sometimes over timescales as short as a few days, implying a significant portion of the radiation is being emitted from a small region of the order of light days in size (e.g., Peterson 1997).   Instabilities leading to hot spots in the disk can produce such variability (e.g., Mangalam \& Wiita 1993).

Around 10 percent of AGN are radio-loud, and are characterized by relativistic plasma jets, which are extended linear structures that transport energy and particles from the compact central region out to kpc or even Mpc scale extended regions (e.g., Peterson 1997).  The emission from these jets results from nonthermal processes, primarily synchrotron radiation, due to highly relativistic electrons spiraling around magnetic field lines.  Inverse Compton scattering of either these synchrotron photons or thermal ones, perhaps from the accretion disk, off the relativistic electrons produce X-rays and $\gamma$-rays.  AGN are variable in every waveband that has been studied: the accretion disk emission from AGN is predominantly at UV and optical frequencies and therefore arises extremely close to the central engine, whereas emission in other bands, such as radio, gamma-rays and some X-rays, emanate from greater distances, often tens of parsecs, out along the jet and even the optical and UV bands are dominated by the nonthermal jet emission (e.g., Peterson 1997). These jets are often highly collimated and may appear to be either single or twin jets; in the former case, the counter-jet is expected to be present and of nearly the same intrinsic brightness, but it appears much fainter than the obvious jet. This difference is due to Doppler boosting, which enhances the apparent surface brightness of the jet approaching the observer and decreases that of the receding counter-jet (e.g., Urry \& Padovani 1995). Knots within relativistic jets may appear to exhibit superluminal velocities if the jets are viewed at small angles. This Doppler boosting also makes sources pointing close to the line of sight appear more variable and such radio-loud AGN are usually called blazars. Variability from blazars often becomes more pronounced at higher frequencies (e.g., Marscher \& Gear 1985; Marscher \& Travis 1991).  

Measurements of the amplitudes and timescales of the fluctuations from relativistic jets are among the few tools we have for understanding their nature and so modeling the variations arising from radio-loud AGN is very important. Some time ago, Marscher \& Gear (1985) produced simple models that included propagating shocks, assumed to arise from disturbances in the inner portion of the jet, heading downstream along the jet in which both the magnetic field and density of the relativistic electrons decrease.  This model provides for variability from radio through X-ray bands based on the assumption that synchrotron emission dominates.  Observed long-term (month to years) variations in fluxes and polarization at various radio frequencies can be understood within this shock-in-jet picture (e.g., Hughes, Aller \& Aller 1991).  To look at faster variations Marscher \& Travis (1991) performed basic numerical simulations of a shock propagating through a medium undergoing hydromagnetic turbulence in a relativistic jet and showed that reasonable light curves could be produced. 
Evidence for the presence of turbulence in relativistic jet flows, particularly upstream of recollimation shocks is now quite strong (e.g., Marscher et al.\  2008).  Recently, Marscher (2014) carefully modeled in detail a region of a turbulent jet with a standing conical shock and produced light curves at different frequencies as well as polarization variations with good resemblance to observations.  Here we also model emission from a limited portion of a jet corresponding to a region of enhanced emission, but focus on investigating how the magnitudes of the turbulent velocities and the nature of the spectrum of the turbulence might affect the observed variability.

Several approaches toward relativistic turbulence in astrophysical systems have been proposed. As an alternative to internal shocks, such as those modeled by Marscher and collaborators, a model by Lazar, Nakar \& Piran (2008) consists of randomly oriented emitters in a relativistically expanding shell; this is designed to provide a mechanism for extremely fast variability from gamma-ray bursts. They employed relativistic subjets, but their paradigm of relativistic emitters does not include a relationship dictating the relativistic velocity across the turbulent energy cascade that is assumed to be present in our model. Narayan \& Kumar (2008) model variability in gamma-ray bursts with a turbulent spectrum more similar to ours, but focus on the relations to timescales, pulses, and variability parameters that are not especially relevant to AGN jets.  We find that Zrake \& MacFadyen's (2013) special relativistic hydrodynamic approach serves as the most useful way to compare a relativistic turbulent spectrum with a standard Kolmogorov turbulence spectrum, since it provides a specific relationship for the Lorentz factor of a turbulent eddy of a given size, given an assumption about the maximum velocity or maximum Lorentz factor of the turbulence. This maximum velocity is a key parameter that is not ab initio constrained by the maximum scale of the turbulence, which, however, cannot plausibly be greater than the diameter of the jet.

Our approach is to use numerical modeling to simulate the variability of the observed fluxes at an arbitrary electromagnetic frequency (nominally optical) arising from turbulent jets and to analyze the variations over time. This work is based on these models for the energy cascade and turbulence spectra for AGN jets, including an implementation of time delays. The paper is structured as follows. In Section 2, we discuss the relativistic and nonrelativistic turbulence spectra we implement in the program, as well as the geometry dictating the time delays. In Section 3, we plot simulated light curves and then calculate simulated power spectral densities (PSDs) corresponding to various input parameters. In Section 4, we analyze the results, briefly compare them with observations, and give our conclusions. 

\section{Methods}
\subsection{Turbulence Spectra}
We begin by modeling the Kolmogorov energy spectrum for a turbulent jet, in which turbulence is characterized by circular eddies turning around and beaming varying emission over time (e.g., Kundu et al.\ 2012). The largest eddy is taken to correspond to the width of the jet at the location of a region of emission enhanced by a shock.  With this standard turbulence spectrum there are relationships dictated between the size, number, and velocity of the smaller eddies that produce the overall spectrum of variations. Of course the Kolmogorov spectrum is derived for non-relativistic turbulence and therefore we ultimately replace the velocity calculations with more accurate relativistic relationships. Nonetheless, this approach provides the only plausible way to couple the size and number of eddies and thus forms a general basis by which the turbulent spectrum can be initialized. We wrote Fortran codes to calculate the theoretical observed fluxes as functions of time based on rest-frame emissions and variable input parameters for bulk velocity of the jet, maximum turbulence velocity in the jet frame, and viewing angle.
\\
\indent In our paradigm, the jet region is comprised of turbulent eddies, with a level of refinement, $i$, ranging from 0 to $k$ (we set $k = 9$); each level contains a number, $N_i$, of eddies of given diameter, $l(i)$, turbulent velocity, $v_t(i)$, and turn-over period, $T_i$. The eddies were initialized to have 
\begin{equation} 
N_i = 2^i 
\end{equation} 
and 
\begin{equation} 
l(i) = l_0 (N_i)^{-1/3}, 
\end{equation} 
where $l_0$ is a constant, the size of the largest eddy, which is assumed to be equal to the diameter of the jet near the location of the reconfinement shock. 
The Kolmogorov spectrum also assumes that the speed of the eddy at the $i^{th}$ level is
\begin{equation} 
v_t(i) = v_{t0} l(i)^{1/3}, 
\end{equation}
where $v_{t0}$ is the speed of the largest eddy, which can be taken to be a free parameter indicative of the strength of the turbulence.
This results in one eddy of the greatest size, velocity, and period, and 512 eddies of the smallest we consider.  In principle one should continue this process to a much larger number $k$, corresponding to ever smaller and slower cells until a dissipation scale is reached.  However, this is not necessary for our purposes since any additional variations produced would be at amplitudes too small to be observed and would also require computing power in excess of that available to us. 
\\
\indent We contrast this Kolmogorov spectrum with the relativistic turbulence spectrum derived by Zrake \& MacFadyen (2013). Though not strictly appropriate for the range of relativistic turbulent velocities we study ($0.1c - 0.9c$), the non-relativistic Kolmogorov theory has been used been used in the earlier work and we will  compare it with the more accurate relativistic turbulence spectrum we now implement. For the relativistic turbulent spectrum, we can still utilize the scaling laws for refinement, number, and size of eddies, given by Eqns.\ (1) and (2) but then replace Eqn.\ (3) by implementing a new relativistic velocity relationship through 
\begin{equation} 
\frac{(\gamma_t-1)^3(\gamma_t+1)}{\gamma_t^2} = \frac{C\epsilon^2l^2}{c^6}, 
\end{equation} 
where $\epsilon$ is the energy injection rate per unit mass and $C$ is a constant analogous to the Kolmogorov constant (Zrake \& MacFadyen 2013). For various trials of a given maximum turbulent velocity $v_{t0}$, we calculated the corresponding energy injection rate $\epsilon$, and then the turbulent velocity can be obtained from its relationship with the Lorentz factor, since $\gamma_t = 1/\sqrt{1-\beta_t^2}$,  where $\beta_t = v_t/c$.
\\
\indent In our calculations we take the physical scale, corresponding to the width of the jet behind the recollimation shock to be a reasonable value of $l \simeq 1$ lt-yr,  and set the speed of light  to unity.  This yields a physical time unit of $l/c \sim 1$ yr.  Each of the timesteps in our code is based on the period of the largest eddy, so that for $v_{t0} = 0.9c$, dt $\simeq$ 1 day. 
At each timestep, the velocity of each eddy was calculated, where a randomized initial angular direction for each eddy at each level was employed. This eddy velocity was added relativistically to the bulk velocity of the jet, $v_b$.  Assuming isotropic turbulence, the components of the total velocity in the frame of an observer at rest in a frame parallel to the jet are
\begin{equation} 
v_{tot_x} = \frac{v_{b} + v_{t_x}}{1 + \frac{v_{b}v_{t_x}}{c^2}} ,
\end{equation}
and
\begin{equation}  
v_{tot_y} = \frac{v_{t_y}}{\gamma_{v_b}(1+\frac{v_{b} v_{t_x}}{c^2})}, 
\end{equation}
where $v_t$ is the turbulent cell's velocity, and $x$ is along the jet and $y$ is perpendicular to it.
\\
If $\theta$ is the viewing angle between the observer's line of sight and the bulk flow of the jet, then the total angle between the line of sight and the location of the emission from the turbulent eddy is  given by
\begin{equation}
\theta_{tot} = \theta + {\rm arctan}(v_{tot_y}/v_{tot_x}).
\end{equation}
One simplification of our work that should be noted is that we are not incorporating an additional random variable for each turbulent eddy for the azimuthal angle of the eddy's velocity and then taking only a projection of that three-dimensional turbulent velocity along the plane defined by the jet and the line of sight to be used in Eqns.\ (5)$-$(7).    Doing so would make no significant difference to the resulting light curves and power spectra but would substantially increase the computational complexity.

\subsection{Relativistic Beaming}
The total velocity from Eqns.\ (5) and (6) and the angle from Eqn.\ (7) were used to calculate the individual Doppler factor, $\delta$, for each eddy in order to find the observed flux of each one. For simplicity and to stress the effects of the turbulence, the rest-frame emissivity is assumed constant, so the flux of each eddy at each level of refinement $i$ is taken to be equal to the flux of the largest eddy, of $i_{min} = 0$ (which is normalized to 1.0), divided by the ratio of the volume of the largest eddy to the given eddy size, so that,  
\begin{equation} S_e(i) = \frac{S_e(0)}{l(0)^3/l(i)^3}. 
\end{equation}
Now with $S_e$ representing the emitted flux from any eddy, we define $S_o$ as the observed flux from that eddy, where 
\begin{equation} 
S_o = S_e \delta^{\alpha + m}. 
\end{equation} 
Here $\alpha$ is the slope of the synchrotron spectrum, taken as 0.5 for a jet, and $m = 2.0$ for continuous emission from a jet, whereas $m=3.0$ for a shock (Blandford \& Rees 1978), and the Doppler boosting factor is defined as 
\begin{equation} 
\delta=\frac{1}{\gamma_{tot}(1-\beta_{tot}~\rm{cos}~\theta_{tot})} 
\end{equation} 
where $v_{tot}$ and $\theta_{tot}$ respectively represent the total velocity, and viewing angle to the eddy within the jet.
\\
\indent Finally, we summed over the flux from each eddy at each timestep, and this total observed flux per unit time was plotted for varying bulk observation angle $\theta$, bulk velocity $v_b$ of the jet, and maximum turbulent velocity $v_t$ of the largest eddy. We computed results for at least an array of 27 different values: $\theta =$  $3\,^{\circ}\mathrm,$ $10\,^{\circ}\mathrm,$ $30\,^{\circ}\mathrm,$ $v_b =$ 0.9c, 0.99c, and 0.999c, and $v_{t0} =$ 0.1c, 0.3c, and 0.9c, and sometimes for 0.6c as well.

\subsection{Time Delays}
To this basic picture, we added time delays to account for the fact that, as the eddies turn around, there is a greater distance between the observer and the emission being beamed that must be traveled. The differential distance, $x$, along the line of sight is illustrated in Fig.\ 1. The delay is implemented so that, for any eddy at any given moment, the delay in time is given by  $t_{delay} = x/c$  where 
\begin{equation} 
x = 2 ~{\rm sin}(\frac{\theta_e}{2}) ~{\rm cos}(\frac{\pi - \theta_e}{2}) l_o, 
\end{equation} 
and $\theta_e$ is the angle between the line of sight and the location of the emission centroid of the eddy.

The remainder of the time delay can be found by considering the assorted locations of each eddy relative to the center of the largest, thereby placing them at different distances from the observer.  As a simple and reasonable, but certainly arbitrary, way of correcting for this we have organized the eddies into a grid of positions relative to one another; the grid we have chosen is shown in Fig.\ 2 for the first three levels of refinement.  For reasonable simplicity we assume the eddies are positioned so that each of their individual centers is at a specific location within the largest eddy. 
The variable angular location of each eddy is given by 
\begin{equation} 
s = \frac{2\pi/N_i}{i+1} + (k-1)(2\pi/N_i) ,
\end{equation} 
where, again, $N_i$ is the number of eddies of each level of refinement, $i$, and now $k$ is the index identifying the individual eddy at that level, which runs from 1 to $N_i$. This yields a grid where the centers of the eddies are separated at equal angles relative to one another.  The eddies of different sizes are also offset to avoid overlapping.  For example, there are two eddies of $i=1$, chosen to have centers placed at $90^{\circ}\mathrm,$ and $270^{\circ}\mathrm,$ and there are four eddies of $i=2$, at $30^{\circ}\mathrm,$ $120^{\circ}\mathrm,$ $210^{\circ}\mathrm,$ and $300^{\circ}\mathrm.$ The simulated position varies by level of refinement and for each eddy within the level is added to the location of the largest eddy, of $i_{min} = 0$, as 
\begin{equation} 
x(i) = l(0) + l(i)\sin(s) - l(i). 
\end{equation} 
This relative position is added to the distance $x$ given in Eqn.\ (11) to account for the total time delay. 

\section{Results}

\subsection{Light Curves}
Illustrative light curves are presented in Fig.\ 3, showing differences arising from varying the viewing angle, bulk velocity and turbulent velocity, all for the case of the relativistic turbulent spectrum.  The plots for the Kolmogorov spectrum are very similar and so only one is displayed in Fig.\ 4, along with the relativistic turbulence light curve for the same parameters. 
The parameters we have investigated usually produce the fractional variations in observed fluxes between 10\% and 40\%; these are not atypical for observed optical variations from blazars over short to long terms (e.g., Stalin et al.\ 2004; MacLeod et al.\ 2012).

The initial rise and terminal fall of the simulated light curves are transients arising from our implementation of the time delays discussed in Section 2.3 and so we excise these portions of the light curves before computing power spectra.
Much larger Doppler boosting is seen for smaller angles, as illustrated in the top two panels of Fig.\ 3. This illustrates the well known very strong dependence of flux upon viewing angle that arises from Eqns.\ (10) and (11).

At a fixed viewing angle, the observed  flux rises with increasing $v_b$, as shown in the middle panel of Fig.\ 3, but does not vary much when the Doppler factor becomes $> 1/\theta$.  For example, at $\theta = 10^{\circ}$ and $v_t = 0.3c$ the observed flux rises by nearly a factor of three as $v_b$ goes from $0.9 c$ to $0.99 c$ but barely increases as $v_t$ rises further to $0.999 c$ (Fig.\ 3, third panel).  While at $\theta = 30^{\circ}$ there is hardly any change in the flux range or maximum flux with variations in $v_b$ and at $\theta = 3^{\circ}$ the flux rises very quickly as $v_b$ increases from $0.9 c$ to $0.999 c$.   
 
 For fixed angle and bulk velocity, the observed flux is much higher for larger values of the maximum value of the turbulent velocity, $v_t$, as illustrated in the last two panels of Fig.\ 3.  The light curves are also much more skewed toward  flares as opposed to  dips as the value of the maximum $v_t$ rises.  Observational light curves do not show such skewed structures, so it appears unlikely that extremely strong relativistic turbulence is present in blazar jets.

\subsection{Power Spectral Densities}
To characterize the type of variability produced by these model turbulent jets, we found their PSDs, which we computed primarily using an original code and confirmed with the PowerSpectralDensity function in {\it Mathematica}. Prior to performing the calculation, we truncate the initial and final transient portions of the light curves, so as to remove the rise and fall due to time delays; this allows us to compute more accurate PSDs for the typical time range in between.  A few of these PSDs are displayed in Figs.\ 5 and 6; these figures show the default (Dirichlet) and Parzen window functions, respectively. 

The observational PSDs of AGN in both X-rays and optical bands are usually well-fit by a red-noise power-law relationship, at least at lower frequencies, and at least for the lengths of time that observations can be made (e.g., Markowitz et al.\ 2003; Vaughan et al.\ 2005; Gaur et al.\ 2010; Mushotzky et al.\ 2011; Wehrle et al.\ 2013; Edelson et al.\ 2013; Revalski et al.\ 2014).  Therefore by plotting the PSDs from our model light curves on a logarithmic scale, it is easy to observe the negative slope in the low frequency range before a break occurs when the slope flattens into white noise for the higher frequencies. To be more precise, we calculate the break frequency by fitting a slope using a least-squares method, starting at high frequencies.  We increase the range over which the slope is calculated into progressively lower frequencies until a non-zero  slope is measured, indicating the break frequency, which occurs when the PSD transitions from red-noise at low-frequencies to white noise at higher frequencies. Once this break frequency is obtained, we can find the slope of the PSD between the lowest frequency limit and the break frequency to obtain the red-noise power-law slopes we now discuss. 

Table 1 presents the mean PSD slopes for each distinct case, which are effectively defined only by the maximum turbulent velocity and the type of spectrum assumed for the turbulence. The slopes were calculated out to a break frequency of 0.1 in our units, which was at, or slightly below, the nominal break frequency for all cases.  Taking a nominal value of $l = 1$ to correspond to a physical jet diameter size of $\simeq$ 1 light-year provides a fundamental timestep of $\simeq 9$ days for $v_0 = 0.1c,$.  We  computed the PSDs using 3 times this timestep to simplify the calculations, thus the highest measurable frequency is approximately 0.7 $\mu$Hz for this case, taking into account the Nyquist theorem. As approximately 3000 of the expanded timesteps were used for these light curves, the lowest frequency computed for these parameters would be  $1.4 \times 10^{-10}$ Hz.  As seen from a comparison of Figs.\ 5 and 6, and as noted in Table 1, the slopes determined by examining the PSDs computed with the Parzen window are typically modestly steeper than those using a Dirichlet (simple box) window.

Higher velocities or smaller jets correspond to shorter timesteps and correspondingly higher frequencies when scaling these simulations to jets of different sizes. Given a timestep defined such that, for the largest eddy at the lowest level of refinement, with $v_{t} = 0.9c$, dt $\simeq$ 1 day, then for $v_{t} = 0.3c$ and $v_{t} = 0.1c$, the timesteps are 3 and 9 days, respectively.  Since the timestep is based on the period of the largest eddy ($l \sim 1$ lt-yr), smaller jets of half this diameter, for example, would produce a timestep reduced by one-half as well. Using 3300 timesteps, and accounting for a shift towards later times due to the addition of time delays, this results in a maximum range of just over 4000 days, as displayed for $v_t = 0.9c$ in  the last panel of Fig.\ 3).   For $v_{t0} = 0.3c$, using the non-relativistic formulation (in which $v_{t0} = v_t$ for the largest eddy, given by Eqn.\ 3), this results in a factor of 3$\times$ the number of timesteps as for $v_t = 0.9c$ (since the timestep is 3 days as opposed to 1).  When calculating the timetstep based on the velocities determined by the more complicated Eqn.\ (4), there is a less direct correspondence with increasing $v_{t0}$, but of course the maximum time range for the smaller velocities increases from the standard 3300 days set by $v_t = 0.9c$.  The last two panels of Fig.\ 3 illustrate how the time range increases as the turbulent velocity decreases.

\section{Discussion and Conclusions}

An examination of all the light curves, in addition to the samples given in Figs.\ 3 and 4, reveals that at fixed $\theta$ and for $v_b$  at least $0.9 c$, the variations induced by turbulence are typically in the several percent though tens of percents.  These are rather typical magnitudes for blazar fluctuations over timescales of weeks or greater.  We find that the effects of different values of the maximum turbulence speed are small as long as $v_t \le 0.3 c$. These light curves are similar to those seen in actual data, which typically show fluctuations that are roughly symmetrical about a mean flux level, or only modestly skewed upward.  However, for very strong turbulence ($v_t = 0.9 c$), our models yield many large, fast flares, producing light curves that are extremely skewed toward high values.  Such high and skewed flux variations are very unusual in actual observations and so we can conclude that such high maximum relativistic turbulent velocities are not likely to be present very often.

Our analysis of the PSDs shows that the slopes of these relativistic turbulent jet models do match the range generally observed in radio-loud AGN. Typical PSD slopes for optical emission from radio-loud AGN, determined from recently produced and lengthy Kepler satellite data sets, range from $-1.4$ to $-2.2$ (Wehrle et al.\ 2013; Edelson et al.\ 2013; Revalski et al.\ 2014), but are most often between $-1.7$ and $-2.1$. 
Some much steeper slopes, ranging from $-2.6$ to $-3.3$, have been calculated from Kepler data on Seyfert galaxies  (Mushotzky et al.\ 2011); however, somewhat shallower slopes have been found in a reanalysis of some of this data that allowed for combining results to more extended periods (Carini \& Ryle 2012). Even if correct, these steeper slopes presumably correspond to emission variations arising from accretion disks in Seyfert galaxies, not jets in blazars, and therefore are not relevant to the radio-loud AGN we are considering. 

All the simulated PSD slopes we found are in the range $-1.70$ to $-2.15$ and hence do agree with the great majority of the observations. While this is most encouraging, it is by no means a confirmation of this jet turbulence scenario in that similar slopes can be obtained from other models, including multiple flares on accretion disks (e.g., Zhang \& Bao 1991; Mangalam \& Wiita 1993), although those models are not believed to be important for radio-loud AGN.

We now note the key results from these calculated PSD slopes. First, the slope of the PSDs depends significantly on the turbulent velocity, but hardly at all on the  bulk velocity and only weakly  on the viewing angle, which is why we present mean values of the slopes in Table 1.   (Of course the bulk velocity and viewing angle strongly affect the magnitude of the fluctuations, and hence, the magnitudes of the PSD, as well.)  Table 1 shows that this slope becomes slightly shallower as the turbulent velocity increases for  the  relativistic (Zrake \& MacFadyen 2013) turbulent spectrum, though there is no difference in the means for $v_t$ values of $0.6c$ and $0.9c$.  However, the slopes obtained for our random realizations at $v_t = 0.9$ have a much greater dispersion and, while they are still independent of $v_b$, they tend to be steeper at smaller viewing angles.  We can understand this because the light curves for very high turbulent velocities are far more asymmetrical, with many large upward excursions.  Even a few such very large fluctuations can noticeably modify the PSD slopes over the lengths of time for which our simulations could be run.

We can compare the differences between slopes of the PSDs calculated according to the relativistic and nonrelativistic turbulence spectra. The relativistic turbulent spectrum PSDs are steeper than the nonrelativistic ones when all parameters are the same, except for $v_t = 0.6c$.  That value for  the Kolmogorov spectrum is also an outlier, in that its magnitude exceeds the average slope calculated for both the $v_t$ values of $0.1c$ and $0.3c$.  This is likely due to our standardization of the break frequency at 0.1 for the tabulated results, as the basic trend toward shallower slopes at larger turbulent speeds is better observed when the PSD is found using individualized break frequencies for different parameter sets. With this exception, we note smaller differences between the relativistic and non-relativistic PSD slopes are found as the maximum turbulent speed increases.   We note that this could arise because smaller differences exist between the velocities found for the nonrelativistic turbulence spectra relative to the relativistic case for increasing levels of refinement.   As the differences in slopes found between the Kolmogorov and Zrake \& MacFadyen (2103) turbulence spectra are modest, it does not seem as if the data can distinguish between them.

In summary, we analyzed the PSD  for each case we computed, and found that only the turbulent velocity significantly affects their slopes. This follows logically due to the assumptions that the bulk velocity of the jet and the viewing angle are assumed to be time-independent, whereas the turbulent velocities vary over the spectrum of eddies, each of which are each positioned at effectively random angles in our models. According to our model based on a relativistic turbulence spectrum, and accounting for the modest effects of time delays, we obtain PSD slopes within a relatively limited range of $-1.92$ to $-2.15$, corresponding to maximum turbulent velocities in the broad range between $0.9c$ and $0.1c$. These values fall within the range ($-1.4$ to $-2.1$) typically observed for PSDs from the observations of radio-loud AGN.  As the steeper portion of this PSD slope range is more commonly observed our results support the hypotheses that turbulent jets can explain the observed variations.  In addition, the shapes of the simulated light curves  at lower turbulent velocities are generally similar to observed ones; however, we can conclude  that extremely high turbulent velocities should not be common.

\newpage
\begin{center}
Table 1: PSD slopes\\
\begin{tabular}{cccc}
\hline
$v_t$ &  Kolmogorov & relativistic (Dirichlet) & relativistic (Parzen)\\
\hline
0.1c & $-1.90 \pm 0.003$ & $-2.09 \pm 0.003$ & $-2.15 \pm 0.001$\\ 
0.3c & $-1.86 \pm 0.012$ & $-2.03 \pm 0.006$ & $-2.14 \pm 0.005$\\
0.6c & $-1.95 \pm 0.030$ & $-1.92 \pm 0.014$ & $-2.05 \pm 0.006$\\
0.9c & $-1.83 \pm 0.097$ & $-1.92 \pm 0.094$ & $-2.05 \pm 0.100$\\
\hline
\end{tabular}
\end{center}

\newpage
\begin{figure}
\begin{center} \includegraphics[scale=0.3]{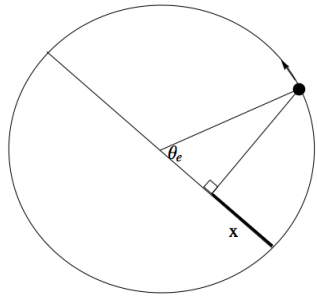} 
\caption{The circle represents a sample eddy, $\theta_e$ is the varying angle made with the line of sight as the eddy turns around, the line across the diameter represents the line of sight, and the small black circle is the zone from which that eddy's radiation is taken to emerge, heading in the direction of the arrow.} \end{center}
\end{figure}

\newpage
\begin{figure}
\begin{center} \includegraphics[scale=0.3]{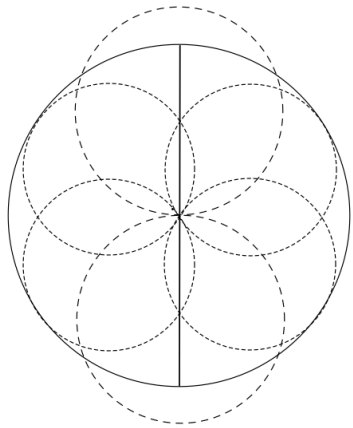} 
\caption{The large circle represents the largest eddy, where $l=1.0$, and the smaller circles show how two subsequent smaller sets of sample eddies are positioned within it. The line across the diameter is the line of sight, and equations (11) -- (13) allow us to calculate the $x$ component along the line of sight for each individual eddy.} \end{center}
\end{figure}

\newpage
\begin{figure}
\begin{center}
\subfigure {
\label{fig:first}
\includegraphics[scale=0.3]{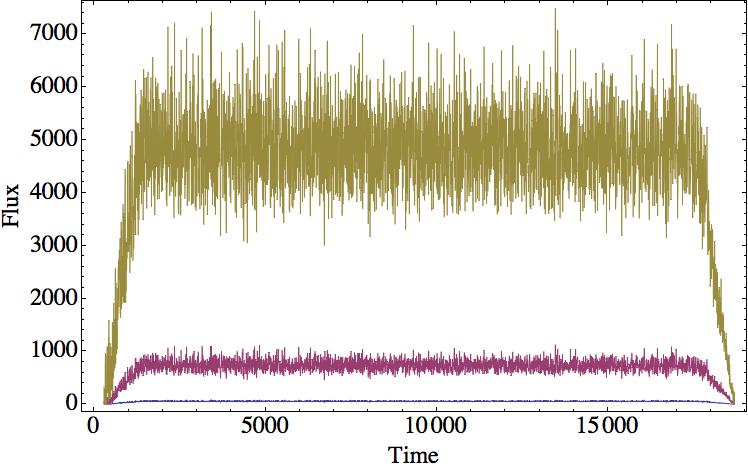} }
\subfigure {
\label{fig:second}
\includegraphics[scale=0.3]{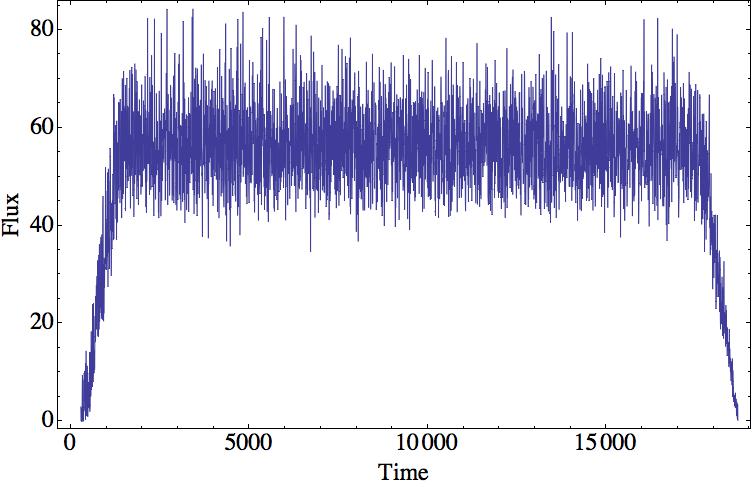} }
\subfigure{
\label{fig:third}
\includegraphics[scale=0.3]{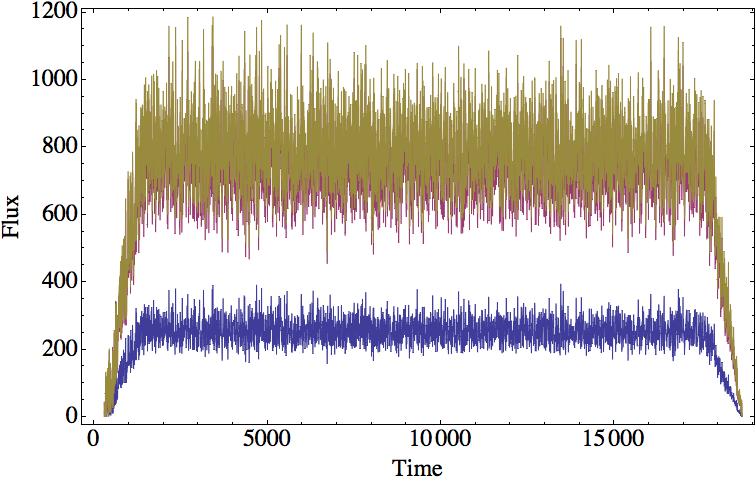} }
\subfigure{
\label{fig:fourth}
\includegraphics[scale=0.3]{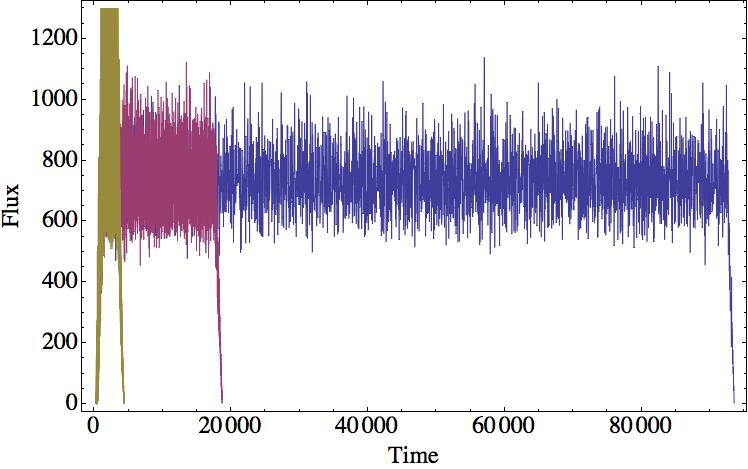} }
\subfigure {
\label{fig:fifth}
\includegraphics[scale=0.3]{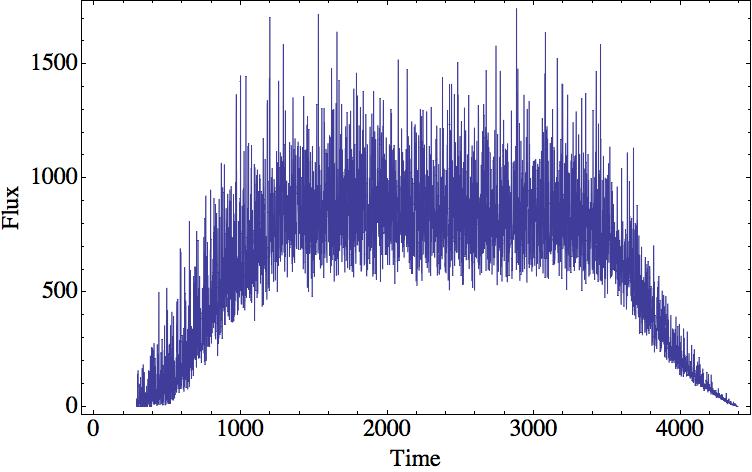} }
\caption{Models of observed fluxes corresponding to variations based on one parameter; the other parameters remain constant at the default values of $10^{\circ}\mathrm,$ $v_b = 0.99c$, and $v_t = 0.3c$. Top left panel: Light curves represent viewing angles of $3^{\circ}\mathrm,$  $10^{\circ}\mathrm,$ and $30^{\circ}\mathrm,$  from top to bottom.  Top right panel: The light curve for $30^{\circ}\mathrm,$ plotted separately to show the scale of fluctuations.  Middle panel: Light curves corresponding to  bulk velocities of 0.999c, 0.99c, and 0.9c, respectively, from top to bottom. Bottom left panel: The light curve for $v_t = 0.9c$ turbulence has the highest amplitude but is computed for the least time (though for an equal number of orbital periods, due to the faster velocity), whereas $v_t = 0.3c$ is the middle case, and the $v_t = 0.1c$ is the longest light curve. Bottom right panel: The light curve corresponding to $v_t = 0.9c$, plotted separately to show the range of fluctuations.} 
\end{center} 
\end{figure}
\label{fig:subfigures}

\newpage
\begin{figure} \begin{center}
\subfigure[Nonrelativistic light curve]{
\label{fig:sixth}
\includegraphics[scale=0.25]{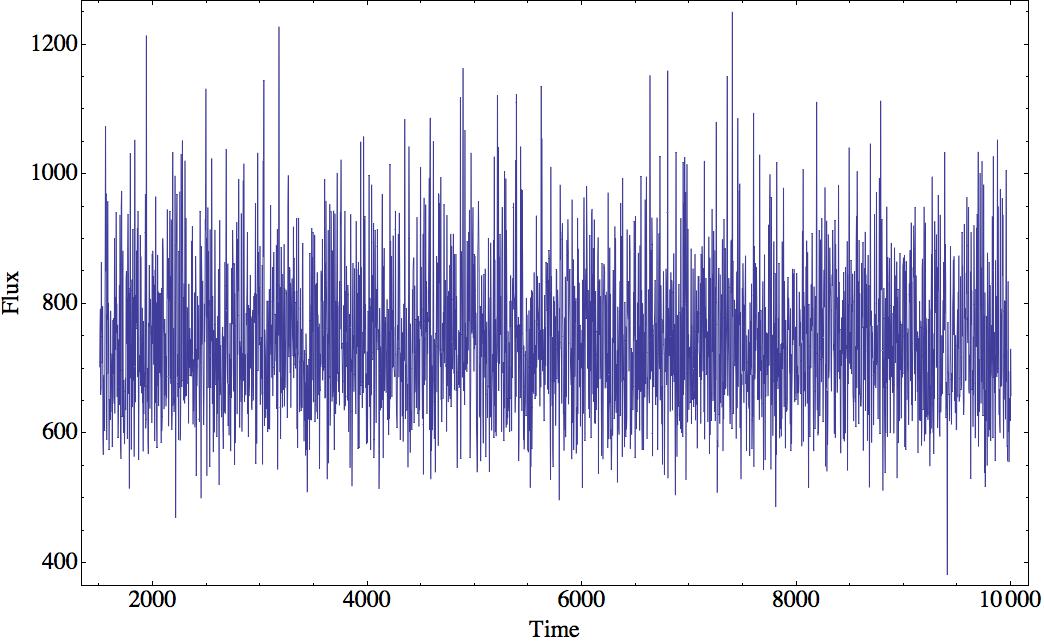} }
\subfigure[Relativistic light curve]{
\label{fig:seventh}
\includegraphics[scale=0.25]{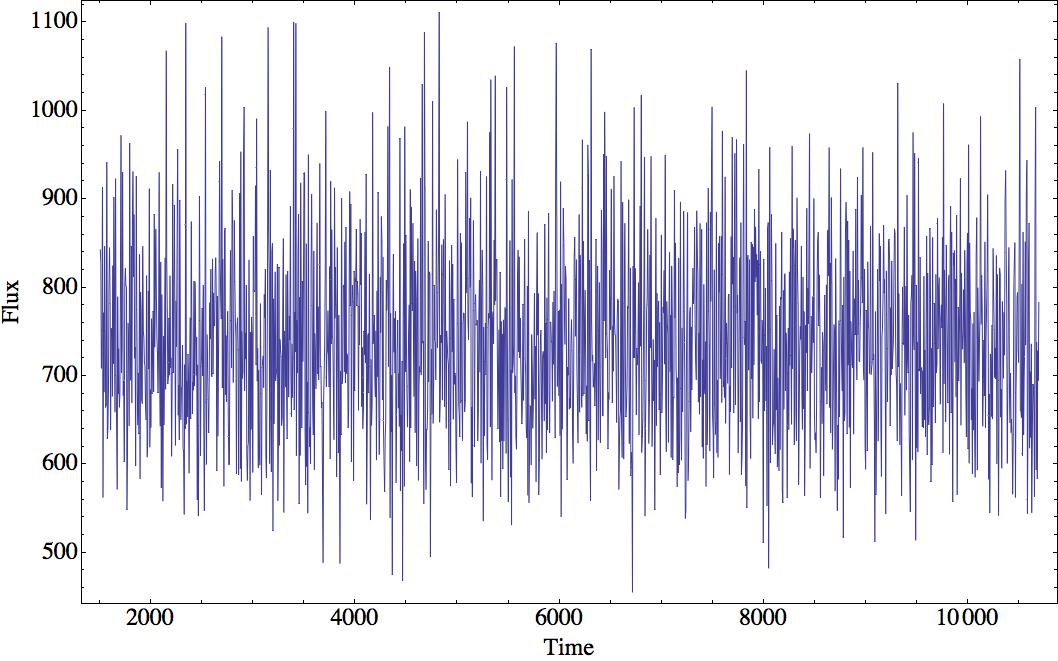} }
\caption{Nonrelativistic (Kolmogorov, left) and relativistic (right) light curves (with time delays truncated) for $10^{\circ}$ viewing angle, 0.99c bulk velocity, and 0.3c turbulent velocity.} 
\end{center} \end{figure}

\newpage
\begin{figure} \begin{center}
\subfigure{
\label{fig:ninth}
 \includegraphics[scale=0.28]{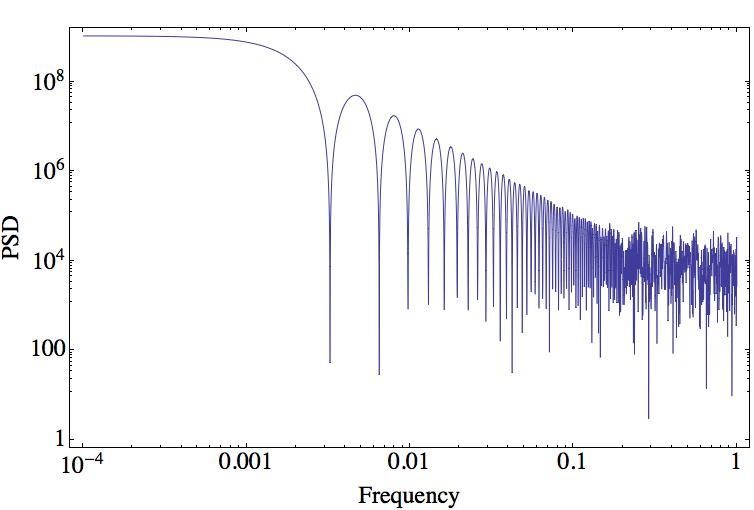}  }
\subfigure{
\label{fig:tenth}
\includegraphics[scale=0.28]{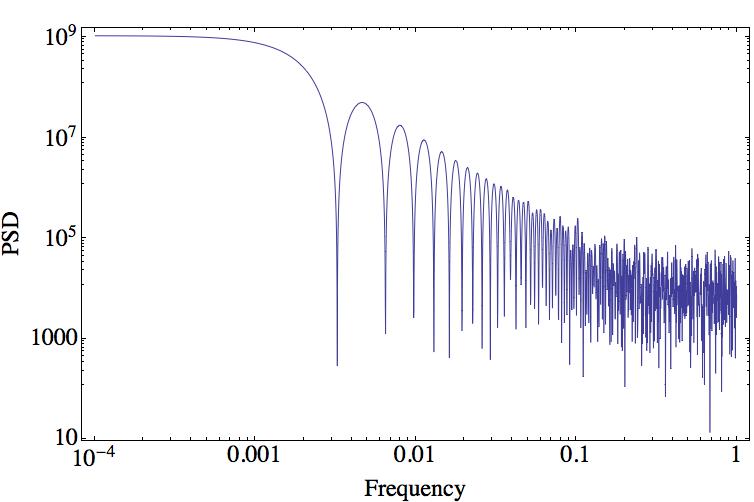} }
\subfigure{
\label{fig:eleventh}
 \includegraphics[scale=0.28]{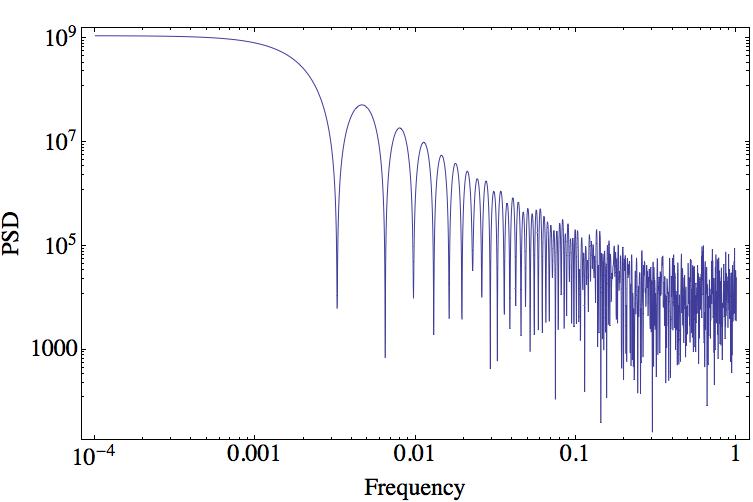} }
\subfigure{
\label{fig:twelvth}
 \includegraphics[scale=0.28]{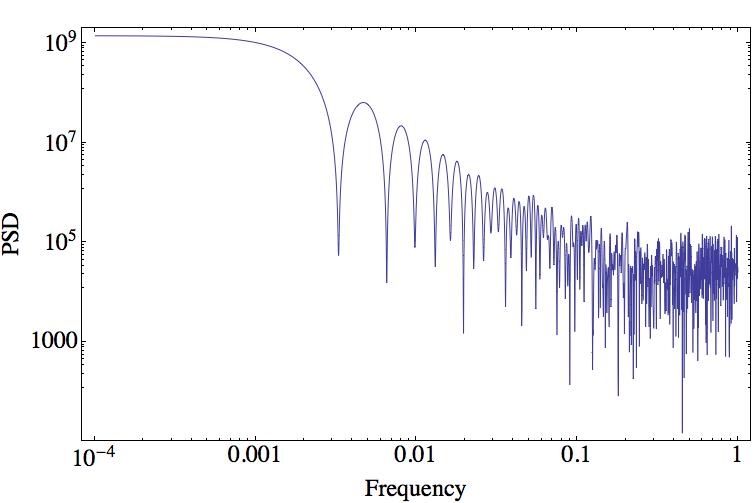} }
\subfigure{
\label{fig:thirteenth}
 \includegraphics[scale=0.28]{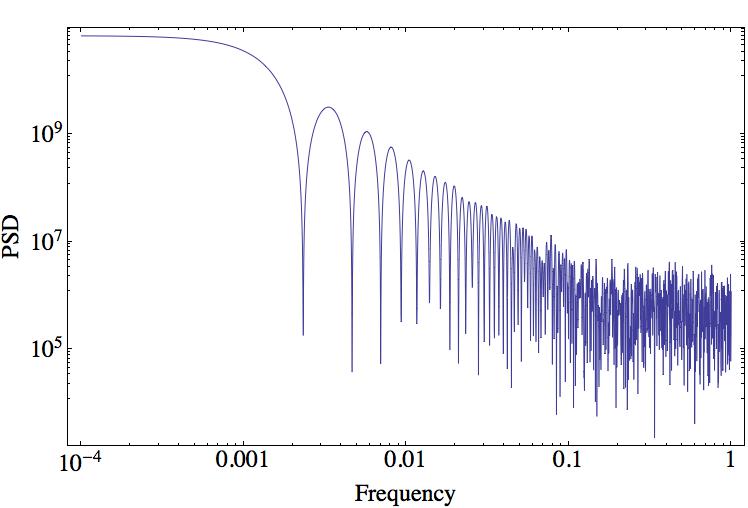} }
\caption{All PSDs correspond to a viewing angle of $10^{\circ}\mathrm,$ and a bulk jet velocity of 0.99c, computed using a simple Dirichlet (box) window function.  From the top and left downwards the turbulent velocities are 0.1, 0.3, 0.6 and 0.9c, respectively, for the relativistic turbulent spectra; the bottom panel is for the non-relativistic turbulent spectrum, with $v_t = 0.3c$.} 
\end{center} \end{figure}
\label{fig:subfigures}

\newpage
\begin{figure} \begin{center}
\subfigure{
\label{fig:ninth}
 \includegraphics[scale=0.28]{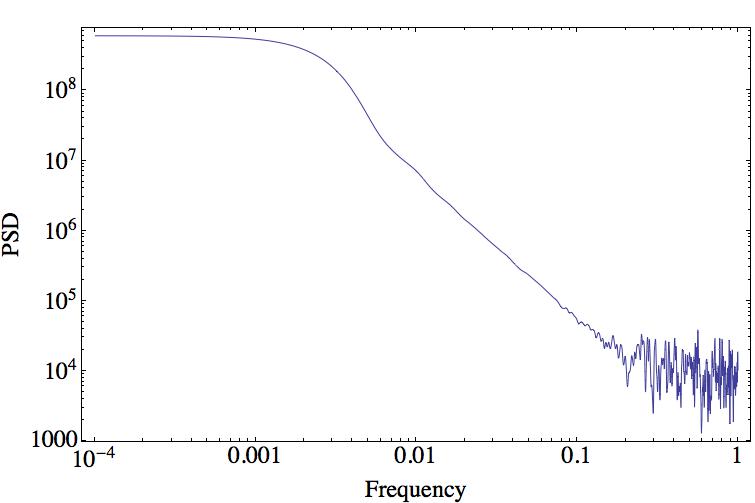}  }
\subfigure{
\label{fig:tenth}
\includegraphics[scale=0.28]{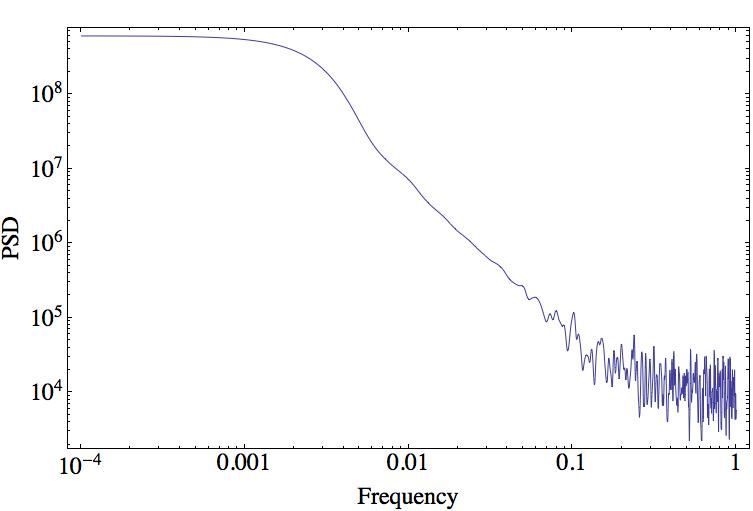} }
\subfigure{
\label{fig:eleventh}
 \includegraphics[scale=0.28]{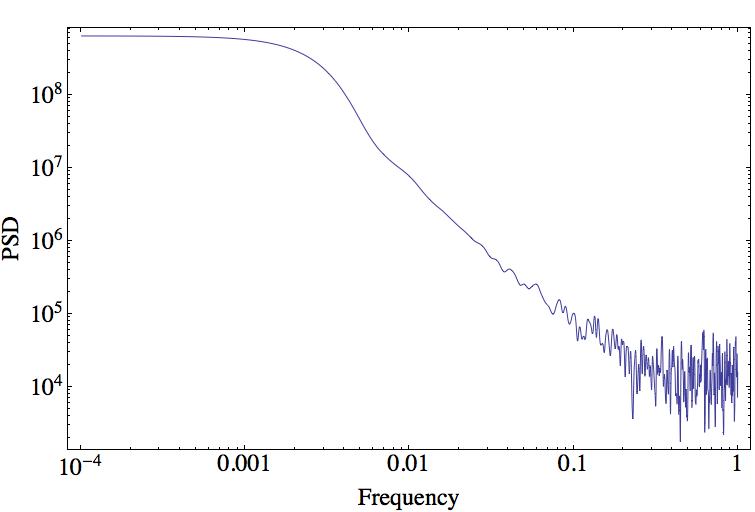} }
\subfigure{
\label{fig:twelvth}
 \includegraphics[scale=0.28]{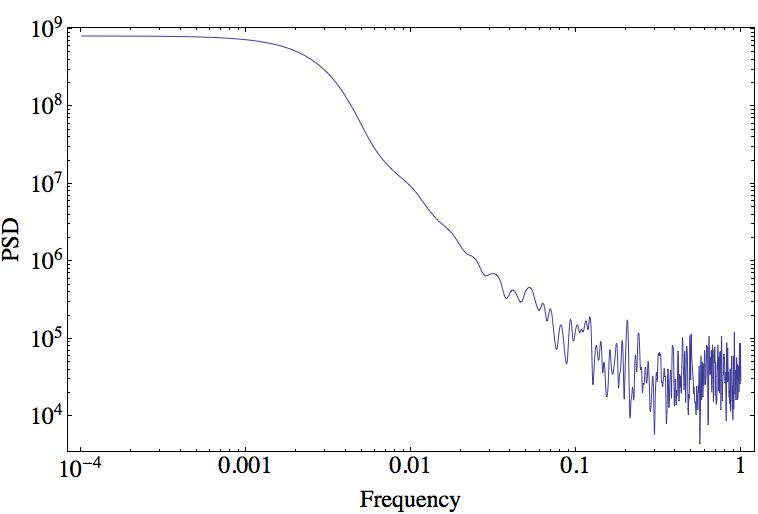} }
\subfigure{
\label{fig:thirteenth}
\includegraphics[scale=0.28]{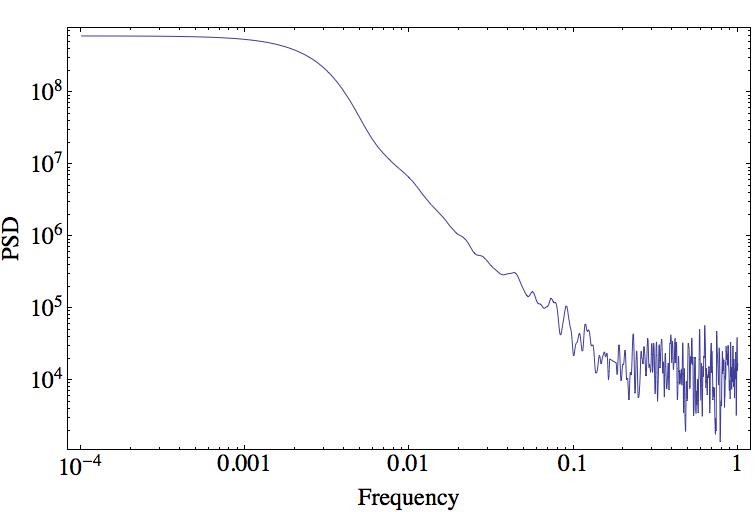} }
\caption{As in Fig.\ 5, computed using a  Parzen window function, which nearly eliminates the ringing seen in Fig.\ 5.}
\end{center} \end{figure}
\label{fig:subfigures}

\end{document}